\documentclass[a4paper,12pt]{article}
\usepackage{axodraw,color,array}
\usepackage{amssymb,url}

\long\def\note#1{\setbox0=\hbox{\large #1}
\ifdim\wd0<0.85\textwidth
\begin{center}\framebox%[\wd0]
{\color{blue}\quad\large #1\quad}\end{center}
\else
\begin{center}\framebox{\parbox{0.9\textwidth}
{\color{blue}\large #1}}\end{center}
\fi
}

\title{\bf Supersymmetry Breaking: constraint on U(1) R-charge}
\author{\bf Pritibhajan Byakti\\
Saha Institute of Nuclear Physics\\
1/AF Bidhan-Nagar, Kolkata 700064, India.}
\date{}

\def\Eqn#1{Eq.\ (\ref{#1})}

\def\3Eqs#1#2#3{Eq.\ (\ref{#1}), (\ref{#2}) and (\ref{#3})}

\begin{document}

\maketitle

\begin{abstract}
Holomorphy of the superpotential promotes any continuous symmetry group $G$ to
a {\sl complexified} symmetry group $G_\mathbb{C}$ of the superpotential
\cite{wess-bagger,Argyres:susy2001.pdf}. For U(1) symmetry this means that
the superpotential is not only  invariant under U(1)  phase rotation but also
under some scaling. We use complexified R-symmetry   to
study the connection between  choices of U(1) R-charges and existence of
runaway directions as well as supersymmetry breaking global minimum in generic
and calculable models.
\end{abstract}

\section{Introduction}
R-symmetries play a crucial role in spontaneous F-term supersymmetry (SUSY)
breaking \cite{Nelson:1993nf,Byakti:2010cm}. It was shown by Nelson and
Seiberg \cite{Nelson:1993nf} that the existence of a U(1) R-symmetry is a
necessary condition and spontaneously  broken U(1) R-symmetry is a sufficient
condition for SUSY breaking in {\sl generic} and {\sl calculable} models.
Generic and calculable models are the dynamical supersymmetry breaking models
whose
low energy theories are described by effective and generic (superpotentials
contain all the terms allowed by symmetries of the theory) supersymmetric
Wess-Zumino (WZ) Lagrangians. In some generic and renormalizable WZ  models, it
is found that there is a connection between existence of runaway directions and 
choice of U(1) R-charges \cite{Ferretti}. Along a runaway direction, scalar
potential keeps on decreasing as we move away from the origin in the field
space.  We will study this issue in generic but effective WZ
models. We will also study the connection of choices of U(1) R-charges with
existence of a supersymmetry breaking global minimum in such theories.

Holomorphy and genericity give us power  to find exact form of
superpotentials \cite{Seiberg:1994bp,Intriligator:1994jr}. Holomorphy of the
superpotential also
promotes any continuous symmetry group $G$ to a {\sl complexified} symmetry
group $G_\mathbb{C}$ of the superpotential
\cite{wess-bagger,Argyres:susy2001.pdf}. For example, if there is a U(1)
symmetry, then superpotential will remain invariant not only under phase
rotation but also under some scaling. Complexified R-symmetry
$U(1)_\mathbb{C}^R$ was used to find connection between U(1) R-charge and
existence of runaway directions in some generic and renormalizable
WZ Models \cite{Ferretti}. As this promotion of U(1) R-symmetry to
$U(1)_\mathbb{C}^R$  has nothing to do with renormalizability
of the superpotentials, we can also use it to find the connection between
existence of runaway directions and choices of U(1) R-charges in generic but
effective theories. After discussing the frame work in sec. \ref{s:frame}, we
study the connection of U(1) R-charges of fields with the 
existence of runaway directions in sec. \ref{s:run} and  with the
existence of SUSY breaking global minimum in sec. \ref{s:susy}.

%%%%%%%%%%%%
\section{Frame-work}\label{s:frame}
We consider generic and effective WZ models with a U(1) R-symmetry ($U(1)_R$)
and some $Z_n$ internal symmetries.  We denote R-charges
of the fields $\phi_i$ are $R(\phi_i)=r_i$ in the normalization $R(\theta)=1$.

Let's assume that some of the fields with non-zero $r_i$ get vacuum
expectation values (VEV) in a vacuum and that one of these fields is $\phi_1$.
We
can re-write the superpotential as follows.
\begin{eqnarray}\label{e:superpot}
 W=\phi_1^{\frac2{r_1}} f(U_2,U_3,\ldots,U_n), \mbox{ with }
U_i=\frac{\phi_i}{\phi_1^{\frac{r_i}{ r_1}}},
\end{eqnarray}
where $U_i$'s and the holomorphic function $f$ are invariants of both the
$U(1)_R$ and $U(1)_\mathbb{C}^R$.  To comment on SUSY breaking, we have to
examine the following F-terms: 
\begin{eqnarray}
 \frac{\partial W}{\partial \phi_1} &=&
\frac2{r_1}\phi_1^{\frac{2}{r_1}-1}\left(
f -\frac{r_i}2 U_i \frac{ \partial  f}{\partial U_i}\right),\\
 \frac{\partial W}{\partial \phi_i} &=& \phi_1^{\frac{2-r_i}{r_1}}\frac{
\partial f}{\partial U_i}  \mbox{ for $i\ge 2$}.
\end{eqnarray}
We see that, if we can solve the following equations
\begin{equation}\label{e:u}
 \frac{\partial f}{ \partial U_i}=0, \mbox{ for $i\ge 2$},
\end{equation}
then all the F-terms except that of the field $\phi_1$ vanish. If we {\sl
assume} that $f$ is also a generic function of $U_i$'s, then each of the above
equations will contain all the variables $U_i$. Therefore, number of independent
equations will never be greater than the number of variables. Hence these
equations can always be solved.  Another important property of the above
equations is that these are invariant under both $U(1)_R$ and
$U(1)_\mathbb{C}^R$. So, we can use the $U(1)_\mathbb{C}^R$ to increase or
decrease $|\phi_1|$ without altering these equations.

 At the solution of these equations, the scalar potential takes the following
form:
\begin{equation}\label{e:kpot}
 V=\left(\frac{\partial^2 K}{\partial \phi_1^\dagger \partial
\phi_1}\right)^{-1} \frac{4}{r_1^2}|\phi_1|^{\frac{4}{r_1}-2} |f|^2,
\end{equation}
and for canonical K\"{a}hler potential it looks simpler:
\begin{equation}\label{e:pot}
 V_{\rm can}=\frac{4}{r_1^2}|\phi_1|^{\frac{4}{r_1}-2} |f|^2.
\end{equation}

Let's now prove why genericity of the superpotentials $W$ do not
necessarily mean that $f$'s are also  generic functions of $U$'s. We consider a
model with four fields $X, Y, Z$  and $\phi_0$ where $R(X)=R(Y)=2$, $R(Z)=1$ and
$R(\phi_0)=0$, and under a $Z_2$ internal symmetry only $Y$ is  odd. We also
consider that the superpotential is non-singular for any finite values of the
fields. Then we can re-write the generic superpotential as follows:
\begin{eqnarray}
 W&=& X h_1(\phi_0) +Y h_2(\phi_0) + Z^2 h_3(\phi_0)\nonumber\\
&=&X \{h_1(\phi_0) + \tilde U_1 h_2(\phi_0) + \tilde U_2 h_3(\phi_0)\},
\end{eqnarray}
where $h_2$ is odd under $Z_2$, $\tilde U_1=\frac YX$ and $\tilde U_2=
\frac{Z^2}{X}$. Clearly $f=h_1(\phi_0) + \tilde U_1 h_2(\phi_0) + \tilde U_2
h_3(\phi_0)$ is not a generic function of $\tilde U$'s because there is no
reason from symmetry ground why the terms like 
$\sum_{i,j\ge2} (\tilde U_1)^i (\tilde U_2)^j h_{ij}(\phi_0)$
where $h_{ij}$'s transform as $(-1)^ih_{ij}$ under $Z_2$, are absent. 
%%%%%%%%%%%%%%
\section{Runaway directions}\label{s:run}
For $r_1\notin[0,2]$, the exponent of $|\phi_1|$ in the \Eqn{e:pot} is negative.
So, the scalar
potential $V_{\rm can}$ monotonically decreases as  $|\phi_1|$ increases and
tends to zero when  $|\phi_1|\to\infty$. Hence  the scalar potential has a
runaway direction if there is a field with R-charge $\notin[0,2]$. If we have
 some U(1) internal symmetries in the theory then definition of R-charges
become arbitrary. For this case, using this arbitrariness if we can make
$r_1\notin[0,2]$, then there will be a runaway direction of the potential. This
demand is true even for non-canonical K\"{a}hler potential
$K(\phi^\dagger,\phi)$, if $\frac{\partial^2 K}{\partial\phi_1^\dagger
\partial\phi_1}$ decreases more slowly than $|\phi_1|^{\frac{4}{r_1}-2}$. 

This result is different from what is obtained by Ferretti \cite{Ferretti}. He
has shown that, in a particular class of generic and renormalizable models,
if there is a field with R-charge not equal to 0,1,2 then there will always be 
a runaway direction. The R-charges which are excluded in our result for
existence of runaway directions are also excluded in Ferretti's result. Some
extra R-charges are also excluded in his case and this might be characteristic
of this class of models.

Now we are going to discuss how our result will change if the theory has some
bigger symmetry, say SU(N), with the help of the famous ISS model
\cite{Intriligator:2006dd}. Fields and  their representations under the global
symmetries of this model are given below:
\begin{center}
 \begin{tabular}{|c|c|c|c|c|c|}
\hline
Field & SU(N) & SU(F)& U(1) & $U(1)_R$\\
\hline
$\Phi$ & {\bf 1} & {\bf Adj + 1 }& 0 &2\\
\hline
$\phi$ & $N$ &$\overline{F}$&1  &0\\
\hline
$\tilde\phi$ & $\overline{N}$ & $F$ & -1  &0\\
\hline
\end{tabular}
\end{center}
where $F > N$.
In this model, K\"{a}hler potential is taken to be canonical and the
superpotential is as follows:
\begin{eqnarray}
 W= h \mbox{Tr} \phi \Phi \tilde\phi -h \mu^2 \mbox{Tr} \Phi
\end{eqnarray}
It is shown in Ref. \cite{Intriligator:2006dd} that there is SUSY breaking.
Now, though we have assigned R-charges of $\phi$ and $\tilde \phi$ to zero,
 using the extra U(1) symmetry we can make them arbitrary as
$R(\phi)=-R(\tilde\phi)=q$. So, this model illustrates the fact that though
R-charges of some fields do not belong to the interval [0,2], yet there is no
runaway direction. However, if we rewrite the ISS superpotential in terms of
SU(N) and SU(F) invariant operators as follows:
\begin{equation}
W= h A  -h \mu^2 B ,
\end{equation}
where $A=\mbox{Tr} \phi \Phi \tilde\phi$ and $B=\mbox{Tr} \Phi$, then we 
see that the operators $A$ and $B$ carry R-charge 2. 

To examine existence of runaway directions in WZ models with bigger symmetries,
we have to rewrite K\"{a}hler potentials and superpotentials in terms of
invariant operators first and then we have to apply the procedure which we have
discussed throughout this section.

\section{Supersymmetry breaking}\label{s:susy}
In this section we will deal with canonical K\"{a}hler potential only.  As
there exists a runaway direction if any field carries R-charge outside
the interval $[0,2]$, there cannot be a SUSY breaking global minimum for
these choices of R-charges. Now the question is, can we get SUSY breaking 
global minimum for all other choices of R-charges? For $r_1=0$, we cannot
re-write the superpotentials in the form given in \Eqn{e:superpot} and we cannot
perform rest of the analysis. For $r_1=2$, the exponent of $|\phi_1|$ in
\Eqn{e:pot} is zero and there may be supersymmetry breaking depending on value
of $f$ and on R-charges of other fields present in the theory. Now, for the case
where $r_1\in ]0,2[$ (i.e. $0< r_1<2$) the exponent of $|\phi_1|$ is positive.
Hence $V_{\rm can}$ decreases if we decrease $|\phi_1|$ using
$U(1)^R_\mathbb{C}$. We can make $|\phi_1|$
arbitrarily small but not zero because then $U_i$'s will be ill-defined. But if
$V_{\rm can}$ is non-zero for
$|\phi_1|=0$, then the scalar potential is discontinuous at $|\phi_1=0|$.

Thus, we have seen that if there is any field whose R-charge is not equal to 0
or 2 then either there is no SUSY breaking or the scalar potential is
discontinuous at the origin of the fields space.

\section{Conclusion}
We have seen that if we assume both $W$ and $f$ are generic functions, then we
can solve all the F-term equations except the F-term equation corresponding to
$\phi_1$ and we can write the scalar potential in a simpler form. On the other
hand holomorphy of the superpotential promotes U(1) R-symmetry to a complexified
U(1) R-symmetry of the superpotential and this enhancement of symmetries has
nothing to do with renormalizability of the superpotentials. Using complexified
U(1) R-symmetry  we find that, if the theory has a canonical K\"{a}hler
potential and some U(1) and $Z_n$ internal symmetries, then  the scalar
potential has a runaway direction if there is a field with R-charge
$\notin[0,2]$. So there is no SUSY breaking global minimum for these choices of
fields. We have also found that if there is a field belongs to the
open interval $]0,2[$ then either
there is no SUSY breaking or scalar potential is discontinuous at the origin of
the fields space.
%%%%%%%%%%
\paragraph{Acknowledgments: }We thank  Palash B Pal for discussions and valuable
suggestions.

\end{document}